\documentclass[twocolumn,showpacs,preprintnumbers,amsmath,amssymb]{revtex4}
\usepackage{graphicx}
\usepackage{dcolumn}
\usepackage{bm}
\begin{document} 
\title{The idea of vortex energy}
\author{V.E. Shapiro}  
\email{vshapiro@triumf.ca}
\begin{abstract}  
This work formulates and gives grounds for general principles 
and theorems that question the energy function doctrine and its 
quantum version 
as a  genuine law of nature without borders of adequacy. 
The emphasis is on the  domain where the energy of systems is 
conserved -- I argue that only in its  tiny part the energy 
is in the kinetic, potential and thermal forms describable  by a 
generalized thermodynamic potential, whereas otherwise the 
conserved energy constitutes a whole 
linked to vortex forces,  and can be a factor of things like 
persistent currents and dark matter. 
\end{abstract}
\maketitle  
\subsection*{The motivation and scope}
 The physics  of phenomena is chiefly perceived
through the interactions  given by the energy function
in line with the
principles of holonomic mechanics and equilibrium thermodynamics.
This brilliantly unifying guideline created 
by Euler and Lagrange has found way in all
pores of physics, and interpretations have  spread out as if 
the guideline is a genuine universal law of nature.  
 But with no outlined borders of adequacy,  the law is 
 a  default belief, a source of  circular theories  and
fallacies.

In this regard, worth recalling the forces called circulatory or 
vortex with all their cumulative impact beyond the energy 
function pattern that can be huge, as exposed since 19th century, 
 e.g. [1,2] and the byword ``dry water'' stuck to 
 viscosity-neglect  hydrodynamic studies as inadequate, see 
 Feynman lectures [3]. Also since 19th century, e.g.  
[4,5], the failure of the pattern was exposed in mechanics and 
other fields due to the reaction forces of ideal non-holonomy, 
performing no work on the system, as is the case of rigid bodies 
rolling without slipping on a  surface.
  Recall  also a general symmetry argument 
provoked by $H$-theorem of Boltzmann and showing the fundamental 
 reversibility Loschmidt's paradox  [6] on the way to conform 
 the real world with the energy function pattern.

The physics of nowadays for all that,
now in line with the quantum mechanics  claimed as more genuine 
than that of classical mechanics, further spreads out the  
conviction  in the genuine energy law with no outlined borders 
of adequacy. And it has a commanding influence
on both fundamental and applied research. This common trend
has various sophisticated possibilities to fall into the  
same trap of circular theories, which in my judgment 
occurs  here and there mainly due to playing with concepts of 
entropy and energy.

Indeed,
``You should call it entropy, for two reasons. In the first place 
your uncertainty function has been used in statistical mechanics 
under that name, so it already has a name. In the second place, 
and more important, no one really knows what entropy really is, 
so in a debate you will always have the advantage''
as John von Neumann remarked on in another connection [7]. 
Curiously enough,  it no less concerns what energy really is. 
This note will try to sketch my approach to it and the
 perspective on the energy perceptions I came to on this way; 
here it is  through 
 formulating and giving ground for relevant general principles 
and theorems. They rectify and develop the idea of energy duality
claimed with insufficient argument in [8] in connection with the 
strong vortex effect of high frequency fields, and point clear  
to the  conserved energy linked to vortex forcing which is  
complementary to all forms of energy function edifice. 

\subsection*{The ``energy cake'' dilemma and equal footing 
theorem}
The established consistent pattern of the world around
is basically relaxation  to roughly recurrent trends.
It implies the ubiquity of  irreversible forces as the 
generalized forces whose infinitesimal work depends on the 
 path  of system  motion rather than just its instant state.
An important mechanism is a manifold   back-reaction of media.

May one then refer the irreversible forcing to the
averaging of irrelevant variables of a conservative many-body 
system given by a microscopic Hamiltonian and random initial
conditions? The answer to this widespread cue is no [8]. 
The cue   
misleads in the question of both statistical and dynamical 
(asymptotic over fast motion) averaging --
there is no way  to come to the irreversible behaviors 
 from the formalism of energy functions unless resort to
the inexact reasoning residing in the  averaging methods and 
truncations irreducible to the separation by canonical 
transformations.

At the same time the perception of myriad of outer influences 
even treated as time-varying Hamiltonian interactions is 
inevitably via smoothing which barely complies with the exact 
separation given by  canonical transformations,  hence, 
contributes to the irreversible  forcing along with 
arbitrariness in  modeling the trends. This is like eat cake and 
have it:
On the one hand,   the irreversible forces, unlike reversible, 
cannot be derived from a Hamiltonian or effective potential.
On the other, insofar as  the true physics of phenomena is 
perceived through the interactions given by energy functions, 
so should be the physics of irreversible phenomena.

The ``energy cake'' dilemma formulated above is inherent 
to the perception and it imports fundamental inexactness in 
 reasoning in terms of energy function.
There is no other way  to account for that but to integrate the 
energy  formalism with a tentative (statistical) measure of 
energy blur/relaxation rates. This element pertains to both 
classical and quantum mechanics descriptions. The uncertainty 
principle of the latter is related to the postulated discreteness 
of energy transfer, has nothing to do with the dilemma,
and the integration in point
puts both descriptions on an equal footing.

Many phenomena in  radiation and rays, superconductivity 
and other fields are  commonly referred to as indescribable 
classically, which might be in one's rights  within 
some specific context; as for unconstrained assertions, it is
to be questioned since  contradicts the above theorem of equal 
footing. The same concerns the ideas of quantum computing claimed 
beyond classical physics, if  their principles appear to be true. 

Naturally, 
each of the two mechanics integrated with the element of  
diffusion/relaxation has a niche where it is handier
depending on interest to discrete or continuous sides of 
phenomena. Even for such irreversible phenomena as highly deep 
cooling of matter by hf resonance fields both ways have 
led to its independent prediction, see [9], and we found  the 
classical way  direct, free of any linkage to the uncertainty 
principle defined by Planck constant. The point is not so much 
that this constant is the same for any  nature of canonically 
conjugated variables, it is that the physics of phenomena can 
be perceived through any self-sufficient construction and that one 
can't see through its wall unless allowing for a dual of the 
formalism with respect to wider frameworks.

\subsection*{The entrainment theorem}
The inexactness of the trends
prescribed by an energy function unfolds generally not only
diffusion-like but also exponentially, able to radically change
the system's state, its stability and fluctuations.
The vortex forces act so. The Gibbsian thermodynamics  and the 
theory of generalized thermodynamic potential [10,2] 
commonly accepted in the study of phase transitions, 
transport through barriers, etc. abstract away from that.
The generalized potential  of a system relaxing in steady 
conditions to a density distribution $\rho_{\mathrm{st}}$ 
connects to it by
\begin{equation}
\rho_{\mathrm{st}}(z)=Ne^{-\Phi(z)}, \quad N^{-1}=
\int  e^{-\Phi}d\Gamma
\end{equation}  
where 
the integral is over the volume  $\Gamma$ of system phase space 
variables $z$ and the reversible motion is on surfaces
\begin{equation}
\Phi(z)=const.
 \end{equation}  
The properties of the system mainly depend then on the local
properties of the minima of $\Phi$. The  analogous approach
to systems under high frequency fields is in terms of the picture
where the hf field looks fixed or its effect is time-averaged.
In all this, Eq.~(1) can be viewed as merely redefining  the 
distribution $\rho_{\mathrm{st}}$ in terms of function $\Phi$,
 which is  suitable for the notion of the entropy of system 
states, whereas taking this function as the energy integral of
reversible motion provides the physical basis of the theory,
but implies rigid constraints.

Commonly, the constraints are reasoned  as detailed balance (of
transition probabilities between each pair of system states
in equilibrium) within the framework of autonomous Fokker-Planck 
equations under natural boundary conditions by means of division 
of the variables and parameters into odd and even with respect  
to  time reversal, with  a  reserve on factors like magnetic 
field. This logic, however, is model-bound and ill-suites 
unsteady conditions. Also it begs a question since detailed 
balance is in leading strings, for the reserve is not universal, 
e.g.,  breaks down in nuclear processes.

A different approach to outlining the overall domain of exactness
was suggested in [8] and will be developed here.
As in general, the basis  is in keeping to invariance 
under transformations of variables. Obviously,  $\Phi(z)$
to be the  integral of reversible motion must be invariant 
under univalent transformations $z\rightarrow Z$, of Jacobian
\begin{equation}
|\det \{\partial Z_k(z,t)/\partial z_i \}|=1
\end{equation} 	
where $i,k$ run all components of $z$ and $Z$, for
then not only  $\rho d\Gamma$ is invariant (being a number) 
but also $d\Gamma$ is.

 The environment as a fluctuation/dissipation  source for 
 the system causes another invariance.
Connecting $\Phi$ to the system's energy function  implies 
scaling this function in terms of environmental-noise energy 
level. The energy scale set so  must vary proportionally with 
the energy function
in arbitrary moving frames $Z=Z(z,t)$ to hold $\Phi$ invariant.
Since the energy function changes in moving frames,
this constraint can hold only for the systems  \emph{entrained} 
-- carried along on the average at any instant
for every system's degree of freedom with the environment 
causing irreversible drift and diffusion.

Also account must be taken of that the limit of weak background 
noise  poses as a structure peculiarity
-- transition to modeling of evolution without regard 
to  diffusion. The entrainment constraint then
keeps its sense as the  weak irreversible-drift limit  
grasped  via the scenarios of motion along the isolated 
paths  in line with d'Alembert-Lagrange variational principle. 
Thereat, however, the  principle  still allows for the 
ideal non-holonomic constraints that violate the 
invariance of  $\Phi(z)$. The invariance therefore
necessitates the domain of  entrainment  free of that,  
termed ideal below.

We have reasoned about $\Phi(z)$ (1), but the reasoning holds
for any one-to-one function of  $\rho_{\mathrm{st}}$.
In unsteady conditions for the   systems describable by  
a time-dependent density distribution $\rho(z,t)$,
the adequacy of energy function formalism
requires the entrainment ideal also.
The arguments used above for the systems of steady 
$\rho_{\mathrm{st}}(z)$ become there applicable with
univalent transformations of $\rho(z,t)$ into 
$t$-independent distribution functions.  

The converse is also true:
the behaviors governed by a dressed Hamiltonian $H(z,t)$ 
imply the entrainment ideal and the existence of
a  density distribution $\rho(z,t)$.
As the  velocity of underlying
motion,  $\dot{z}= \dot{z}(z,t)$, is constrained by 
$\dot{z}=[z,H]$ with $[,]$  a Poisson bracket,  
the divergence $div\hspace{1pt}\dot{z}=div[z,H]= 0$ 
and $div (\dot{z} f)=-[H,f]$ for any smooth $f(z,t)$. 
It implies
 \begin{equation}
\partial \rho/\partial t =
[H,\rho]
\end{equation}  	
which determines $\rho(z,t)$ from a given initial distribution 
and the natural boundary conditions preserving
the normalization and continuity, for all other constraints are 
embodied in $H$. In no way the solution to (4)
ceases to exist as unique, non-negative
 and not normalizable  over the phase space of $z$
where $H(z,t)$ governs the behaviors.
 The entrainment ideal there takes  place since
the solution  turns to $\rho(H)$ in the
interaction picture where $H$ is $t$-independent. This
completes the proof.

Thus, the necessary and sufficient conditions where 
the energy function doctrine is duly adequate to the evolution
described by distribution functions  come down to the 
entrainment ideal.
This theorem lays down the overall domain of energy 
function adequacy sought for. It includes  the
systems  isolated or in thermodynamic equilibrium, 
as well as entrained in steady or unsteady environments
generally of non-uniform temperature or indescribable 
in temperature terms so long as the diffusion, irreversible 
drift and ideal nonholonomy can be neglected.

Remark 1.  
The trend of entrainment ideal can  be deprived of evidential 
force already in the close vicinity of the ideal.
In steady condition this can be
not so much due to extremely long observation times as due to
ideal nonholonomy, for the diffusion and irreversible drift are
then enhanced hugely. To see it, suffice to bear in mind that 
the ideal nonholonomy not only reduces 
the number of degrees of freedom relative to the number 
of generalized coordinates, but also 
gives rise, see [5], to equilibrium  states  and also 
steady-motion states that are not isolated but form  
manifolds  of one or more dimensions and asymmetry 
of secular-equation determinants.

Remark 2. 
The evolution of density distribution  $\rho(z,t)$ of system 
states from a given initial $\rho(z,0)$ gives
by itself no insight into the matter of entrainment
even in steady conditions, unlike, say, their multitude from
various initial distributions. The trend of evolving then
to one and the same shape of  $\rho$ means relaxation with the    
mean (``drift'') irreversible forces a factor. 
For the relaxation to steady motion, such forces are of vortex 
type in $\Gamma$ as their forcing  
toward the steady  motion and against it  differ in sign.
They effect both the steady and the transient shapes of $\rho$.
Thereby  the   energy integral of the motion ceases to exist 
with time if set initially, being disrupted  by the drift 
vortex forces along with diffusion. These forces act generally 
stronger than diffusion,  multiplicatively, 
and cannot be compensated by conservative forces. 
Unstable are   then as minima of $\Phi(z)$ (1) as any 
points shifted from them, and the motion states
can appear quite apart from the minima, in defiance of  
the   theory of phase transitions  based on generalized 
potentials.

\subsection*{The  ``energy - energy function'' dualism 
and the conserved energy linked to  vortex forcing}

Let us refine on concepts.
The issue of energy we are raising  relates to the  
systems of finite degrees of freedom that interact 
with the environment whose
influences of short correlation time are accounted for 
 via the notion of entrainment introduced above.
The systems  are assumed describable  by a smooth 
evolution of the density distribution $\rho(z,t)$ of
phase space  states $z$,
a set  of continuous variables $z=(x,p)$
with the generalized coordinates $x=(x_1,\ldots x_n)$ 
and  conjugated moments $p=(p_1,\ldots p_n)$ of proper 
$n$ taken  in neglect of the constraints breaking the 
energy function formalism;  $z$ may include
countable sets of normal mode amplitudes of waves in 
continuous media.
The smoothness of $\rho$ means
\begin{equation}
\partial \rho/\partial t=-div (v\rho)
\end{equation}  
with $v\rho$  the $2n$-vector flux of
phase fluid  at $z$, $t$. 
Eq.~(5) turns into the evolution equation  of $\rho(z,t)$
under natural boundary conditions with
$v$ treated as  operator  on $\rho(z,t)$
that accounts for all constraints on the phase flows;
in neglect of all nonlocal and retarded constraints, 
 $v$ is generally a $t$-dependent field
divergent in $z$.

For the evolution of $\rho$ modeled
by an equation of form
\begin{equation}
\partial \rho/\partial t=
[H,\rho] + I
\end{equation}  
 where $H=H(z,t)$ is now, unlike in Eq.~(4), an arbitrary 
smooth  function taken for a Hamiltonian, and $I$
embodies  all other  interactions, we have
\[I=-div[(v-\dot{z})\rho]
\]
with $\dot{z}=[z,H]$   the local velocity of Hamiltonian 
phase flow and $I$ a canonical invariant. 
 The invariance of $I$ 
holds as in as off the entrainment ideal. Indeed,
a canonical (univalent) transformation $z\rightarrow Z$ 
implies not only the invariance of $\rho$ and  
Poisson brackets but also the constraint 
$\partial Z(z,t){/}\partial t=[Z,G]$ with $G$ 
a function of $z,t$. Hence, on transforming Eq.~(6) we get
\[(\partial \rho{/}\partial t)_Z=
(\partial \rho{/}\partial t)_z-[G,\rho] \]
where the  term  $[G,\rho]$ is to be united with $[H,\rho]$ 
of (6), for just so  a Hamiltonian is to change. 
Thus, Eq.~(6) in the new variables differs  by its r.h.s. 
changing so
\begin{equation}
[H,\rho]+I\rightarrow [H+G,\rho]+I
\end{equation}	
with $I$ to be held invariant. This completes the proof.
 
 It should be underlined 
that the invariance of $I$  is not unconditional but under
canonical transformations and reflects the fact of 
proceeding in modeling from a Hamiltonian.
Since it is governed by a Hamiltonian only
in the entrainment ideal, only there  $I$ reduces  to 
an invariant Poisson bracket; whereas the irreducibility of 
$I$ this way beyond
the ideal exposes $I$ of Eq.~(6) as the source of 
irreversible, hence,
vortex  forcing that  breaks the  invariance of $I$ 
for any choice of $H(z,t)$.

Let us now turn to the concept of energy within the 
framework under study. At that,
while the $x$, $p$ of $\rho(z,t)$ is a set of phase space 
variables, the principle of virtual work on the system and 
the law of energy conservation, which are to be taken 
as prime as so  the material world is perceived, are 
formulated in terms of isolated paths with $x$ and $p$ 
the functions of  $t$.
Naturally, we treat any conceivable isolated paths as 
abstraction  of the kinetics of $\rho$, so  the integrable 
correspondence  between the two descriptions is to imply on 
the principles of continuity and causality. 
 The  $n$ components
$(v_{n+1}, v_{n+2},\ldots v_{2n})$ of the \emph{actual} 
phase flux at $z$, $t$ act then as
 the generalized force conjugated to $x$ and the 
 scalar product
\begin{equation}
v_{n+1}\delta z_1+ v_{n+2}\delta z_2+\ldots v_{2n}\delta z_n
\end{equation}	
represents the virtual work  on the system
irrespective of whether  this sum  is reducible to the 
variation of a scalar function or not. Accordingly,
for the generalized coordinates taken without 
restricting the generality
 in the geometric conditions not involving time explicitly,
the  density power on the  phase fluid comes down to the 
scalar product
\begin{equation}
(v_{n+1}v_1+\dots  v_{2n}v_n)\rho.
\end{equation}	
In particular, the energy of the  system is conserved 
as long as the integral  
 of density power (9) over the whole phase volume 
holds zero,
\begin{equation}
\int(v_{n+1}v_1+\dots  v_{2n}v_n)\rho d\Gamma=0. 
\end{equation} 
 This criterion bears by itself no relation
to the entrainment ideal and shows up
in both entrained and non-entrained systems and  as 
under steady constraints (autonomous Eq.~(5)) as unsteady.

Where the energy of system is conserved, there its energy 
measure exists in strict sense. So, the criterion (10)
outlines the existence domain of the energy measure.
It includes the whole existence domain of the energy 
measure in the entrainment ideal, which is obviously
 where $v$ is a $t$-independent divergent-free 
 function of $z$, but can extend fairly far beyond 
 it - however  far in principle both in drift and 
 diffusion terms of $v$ 
and whether they are retarded and  $t$-independent or not.

Consider in this light  
the conditions of energy conservation in the systems 
governed by autonomous Eq.~(6).  The branch $I$ there 
acts on a par with  $[H,\rho]$ in keeping the  
circulation and  transformations of conserved energy, 
and this takes place as within as beyond the scope of 
entrainment ideal. This  fact means that the energy 
circulating in conditions beyond the ideal is 
indescribable in terms of  energy function.  
We called it vortex form of energy. 
The energy exchange between degrees of freedom is then 
not via detailed balance,  
involves vortex forcing.    At that, the stationary 
conditions of ideal non-holonomy can be viewed as a 
particular case of such form of energy circulation. 

Thus, unlike the conventional
kinetic and potential energies and the thermal 
energy as the chaotic variant of kinetic energy 
within the generalized thermodynamic potential, the form 
of conserved energy circulating in the non-entrained systems
is irreducible to a scalar function of system states. 
This form of energy is integral -- relegation of  its parts 
to conventional is  not uniquely defined, which shows the 
vortex form of energy as complementary to the energy of 
conventional forms. This is what we call the 
energy - energy function dualism.
It has nothing to do with
the particle - wave dualism in quantum  mechanics. 
The notion of energy quantum levels 
as the quanta of physical substance  related to specific 
system states loses then its strict sense, and 
so does the transfer of energy via energy quanta.  

The ability of vortex forces  to radically, 
cumulatively change the system's 
state, stability and fluctuations, as is the case of 
systems under high frequency fields, and the  ubiquity of  
vortex forcing  also under the restrictions 
of system energy  conservation conveys a  suggestion that 
the vortex form of conserved energy can be a factor of 
persistent currents and also  puzzles like  dark matter. 



\end{document}